\documentclass[%
 aps,prb,reprint,twocolumn,
superscriptaddress,
 amsmath,amssymb,
 aps,
]{revtex4}
\usepackage{soul}
\usepackage[dvipsnames]{xcolor}
\usepackage{graphicx}% Include figure files
\usepackage{verbatim} 
\usepackage{dcolumn}% Align table columns on decimal point
\usepackage{bm}% bold math
\usepackage{enumitem}

\begin{document}

\title{Cu NMR study of lightly doped La$_{2-x}$Sr$_{x}$CuO$_4$}% Force line breaks with \\

\author{Marija Vu\v{c}kovi\'{c}}
% \ead{marija.vuckovic@kbc-zagreb.org}
\address{%
Faculty of Science, University of Zagreb, Zagreb, Croatia}%
\altaffiliation{
Present address: University Hospital Center Zagreb, Zagreb, Croatia 
}
\author{Ana Najev}%
\altaffiliation {Present address: Ericsson Nikola Tesla, Krapinska 45, HR-10000 Zagreb, Croatia}
\address{%
Faculty of Science, University of Zagreb, Zagreb, Croatia}%

\author{Biqiong Yu}%
\address{%
School of Physics and Astronomy, University of Minnesota, Minneapolis, MN, USA}%
\author{Takao Sasagawa}
\address{%
Materials and Structure Laboratory, Institute of Science Tokyo, Kanagawa, 226-8501, Japan}%
\author{Nina Bielinski}
\altaffiliation {%
Present address: Department of Physics, University of Illinois at Urbana-Champaign, Urbana, IL, USA}%
\address{%
School of Physics and Astronomy, University of Minnesota, Minneapolis, MN, USA}%
\author{Neven Bari\v{s}i\'{c}}%
\address{%
Faculty of Science, University of Zagreb, Zagreb, Croatia}%
\address{%
Institute of Solid State Physics, TU Wien, Vienna, Austria}%
\author{Martin Greven}%
\address{%
School of Physics and Astronomy, University of Minnesota, Minneapolis, MN, USA}%

\author{Damjan Pelc}%
\address{%
Faculty of Science, University of Zagreb, Zagreb, Croatia}%
\address{%
School of Physics and Astronomy, University of Minnesota, Minneapolis, MN, USA}%

\author{Miroslav Po\v{z}ek}%
 %\ead{mpozek@phy.hr}
\address{%
Faculty of Science, University of Zagreb, Zagreb, Croatia}%

\begin{abstract}
We present a single crystal Cu NMR study of the cuprate superconductor La$_{2-x}$Sr$_{x}$CuO$_4$ with hole doping levels between $x=2$\% and $8$\%. Measurements with short spin echo times enable us to systematically study the local properties of the electronic spin system in the region of the phase diagram where the material evolves from the insulating antiferromagnetic parent phase to the superconducting state. We find evidence for qualitative changes as the Sr concentration increases through $x=5$\% in both NMR spectra and relaxation times, which we interpret as signatures of a low-temperature transition from a state with disconnected metallic islands to a granular metal with tunneling between the grains. These results provide microscopic insight into the physics of a doped charge-transfer insulator in the presence of quenched substitutional disorder, indicate that a connectivity transition occurs, and demonstrate that complex nanoscale electronic phase separation is ubiquitous in LSCO around the doping level where superconductivity first appears in the phase diagram.  

\end{abstract}

\maketitle

\section{Introduction}

Among the numerous families of cuprate high temperature superconductors, La$_{2-x}$Sr$_x$CuO$_4$ (LSCO) is one of the most extensively studied due to its relative structural simplicity and wide range of attainable hole doping levels. As in other cuprate families, the undoped parent compound La$_2$CuO$_4$ is an antiferromagnetic (AFM) charge-transfer insulator; concentrations of doped holes as small as $p \sim 1\%$ induce a transition into a metallic state with strong short-range spin correlations, which becomes superconducting upon further hole doping  \cite{Julien2003}. It was proposed early on that the physics of lightly doped charge-transfer insulators and the insulator-metal-superconductor transition in this low-doping regime might hold the key to understand cuprate superconductivity \cite{Lee2006,Zaanen1985,Keimer2015}. However, fundamental open questions remain.

\begin{figure}
\centering
\includegraphics[trim={0.3cm 0 -1cm 0} ,clip, width=0.8\linewidth,]{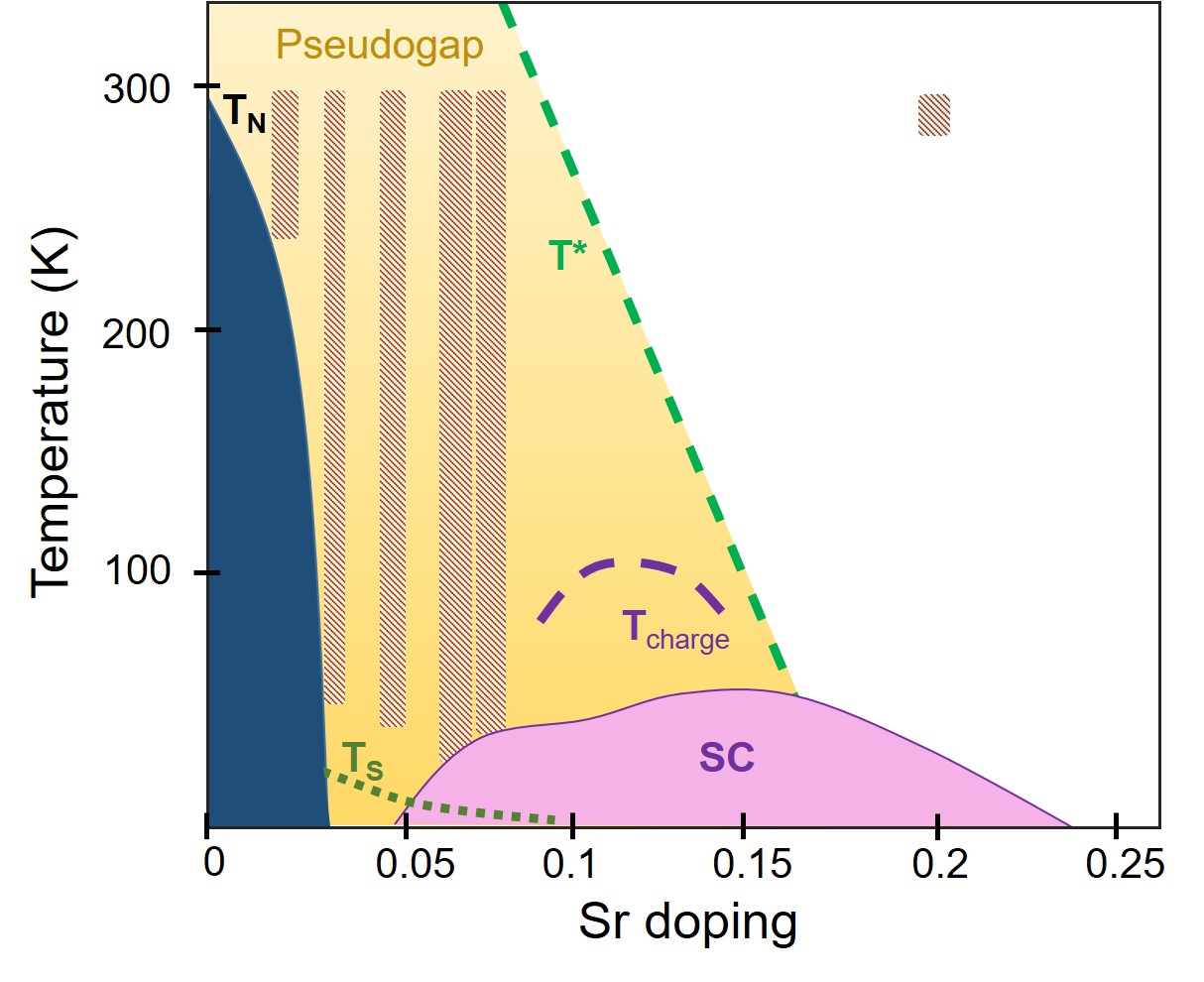}

\caption[A generic LSCO magnetic phase diagram]{Schematic phase diagram of LSCO, based on refs. \citenum{Julien2003} and  \citenum{Keimer2015}. The Neel temperature ($T_N$), spin glass transition ($T_S$), charge order transition ($T_{charge}$) and pseudogap temperature (T*) are shown. Red shaded areas represent samples (doping and temperature ranges)  studied in this work. Previously, Cu NMR measurements have been restricted to dopings above $x$=6$\%$ and temperatures above 100~K \cite{Imai1993, Ohsugi1994, Julien1999}. \label{fazni} }
\end{figure}

Experimentally, both electronic and structural inhomogeneity has been found to be ubiquitous in cuprates \cite{Pan2001, Dagotto2005, Fischer2007, Phillips2003, Fratini2010, Pelc2018, Pelc2019}, and various forms of nanoscale electronic correlated states have been proposed to arise upon the suppression of long-range AFM order (Fig. \ref{fazni}). These include spin glass  \cite{Wakimoto2000, Wakimoto1999, Julien2003, Chou1995, Baek2017}, spin-charge stripes \cite{Imai2017, Imai2021, Braicovich2010}, as well as more complex nematic phases reminiscent of soft matter \cite{Capati2015}. The characteristics of the spin glass phase depend in a complex way upon the details of a given cuprate family \cite{Sasagawa2002}, most importantly the density of point disorder. In particular, in LSCO and other compounds such as YBa$_2$Cu$_3$O$_{6+\delta}$, the presence of a spin glass state has been confirmed in magnetic resonance and muon spin rotation experiments \cite{Julien2003, Stilp2013, Niedermayer1998}. On the other hand, the spin-glass phase seems to be entirely absent in some families where hole doping is achieved though interstitial oxygen atoms far from the Cu-O planes, such as HgBa$_2$CuO$_{4+\delta}$ (Hg1201), at least in the accessible doping range above $p \sim 5$\%. In addition, incommensurate spin modulations consistent with fluctuating stripes have been observed in La-based cuprates using neutron scattering \cite{Wakimoto1999,Wakimoto2000,Fujita2002}. While a controlled theory of charge-doped AFM insulators remains elusive, a tendency toward spatial inhomogeneity has been successfully modeled through theoretical and numerical work \cite{emery1990, Kivelson2003}.

Bulk charge transport measurements in LSCO and other cuprates show metallic behavior above $\sim 50-100$~K, and point to the existence of a Fermi-liquid-like component down to hole doping levels approaching zero  \cite{Ando2004,Li2017}, consistent with ARPES results  \cite{Ando2007}. Remarkably, the effective transport scattering rate of the itinerant subsystem in this regime is nearly doping-independent  \cite{Barisic2019} and similar to the universal rate observed across the cuprate phase diagram  \cite{Barisic2013,Li2016}, despite the dramatic doping-induced changes in the nature and extent of AFM spin correlations. At low temperatures, however, the resistivity of LSCO exhibits an upturn \cite{Ando2004, Ando2001}. The functional form of this upturn undergoes a subtle change around $p = x \sim 5\%$, which points to a possible qualitative change in microscopic properties at this doping, roughly where superconductivity sets in  \cite{Herlach2003HighMF}. Below $x \sim 5\%$, the low-temperature resistivity upturn is exponential and attributed to variable-range hopping  \cite{SeikiKomiya_2009}, which indicates that the $x$ mobile holes localize in cooling toward the spin-glass phase. At doping levels above $x \sim 5\%$, the temperature dependence of the resistivity upturn becomes logarithmic  \cite{Ando1995, Ando2004, Balakirev2003, Wambecq2004, Malinowski2008, Li2016, Komiya2004}, which has been attributed to the formation of a granular metal state \cite{Beloborodov2007,Komiya2004}, in which quantum tunneling of mobile carriers through insulating regions becomes possible. This state, where tunneling effects dominate, persists into the superconducting phase at a microscopic level  \cite{Beloborodov2003, Beloborodov2007}: measurements on La-based cuprates in high magnetic fields show that, once superconductivity is suppressed, a logarithmic upturn remains present deep into the superconducting dome \cite{Barisic2013, Ando1995,Boebinger1996}. It should be noted that the logarithmic upturn in other cuprates appears at a non-universal, family-dependent doping, and was demonstrated to be sensitive to the density of point defects  \cite{Alloul2001,Rullier-Albenque2008, Li2017}. %This is seen, e.g., in slightly Zn co-doped LSCO, where the transition from an exponential to a logarithmic upturn does not occur until the doping level reaches $\sim$ 12$\%$. 

\begin{figure}[t!]
\includegraphics[trim={0 0 0 0},clip, width=0.8\linewidth]{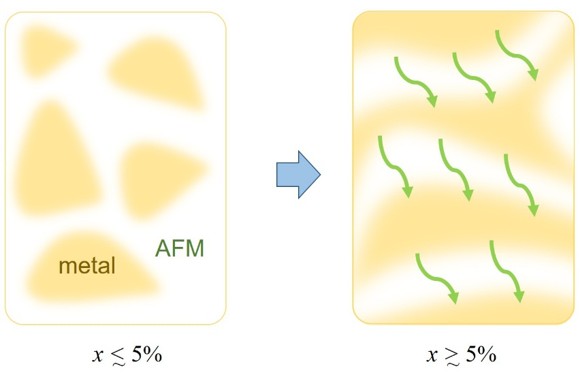}
\caption[Cluster model] {Schematic of possible electronic inhomogeneity for doping levels $x<$5$\%$ and $x>$5$\%$. At low doping, isolated metallic clusters appear. As the clusters grow, tunneling is established, and a cluster metal is formed. A further increase of the carrier concentration does not qualitatively affect the width or dynamics of spectral lines.} \label{cluster}
\end{figure}

Taken together, these features indicate that substantial electronic inhomogeneity and distinct local environments can form in lightly doped cuprates, with the possible appearance of insulating and metallic regions (Fig. \ref{cluster}). Such effects may only be important in cuprates such as LSCO, where significant quenched point disorder induced by substitutional doping masks the underlying behavior of pristine copper-oxygen planes. It is nevertheless essential to obtain deeper understanding, given that all cuprates that can currently be doped below $p \sim 5$\% exhibit similar phenomenology. 

Here we use $^{63,65}$Cu  nuclear magnetic resonance (NMR) as a local electronic probe of LSCO in order to obtain direct microscopic insight into AFM fluctuations, and electronic heterogenieity as the system evolves from the insulating parent compound. NMR is a suitable technique with which to investigate local spin fluctuations and microscopic structures, as it is a local real-space probe that reflects the range and distribution of magnetic fields and electric field gradients in a sample. Furthermore, NMR can provide valuable insight into system dynamics, through measurements of spin-lattice and spin-spin relaxation times. However, being a microscopic probe, NMR can not provide information about the spatial configuration of charges beyond the immediate surroundings of the studied nuclei.

In lightly doped LSCO, detailed NMR experiments have been performed on $^{139}$La nuclei. Cu NMR is a more direct local probe of the quintessential Cu-O planes, but Cu NMR investigations have been limited to temperatures above $\sim 200$~K due to considerable difficulties, such as loss of signal, fast relaxation rates, and wide spectral lines  \cite{Ohsugi1994, Baek2012,Grafe2010,Julien2001,Julien1999, Arsenault2018, Arsenault2020}. In this work, we overcome these difficulties through nuclear spin-echo measurements with short echo times \cite{Imai2017, Pelc2017}, which enable us to measure Cu NMR in a range of doping and temperature that has not been accessed before (Fig. \ref{fazni}). We show that there is a sudden and qualitative change in spectral widths, spin-spin relaxation rates, and their temperature and frequency dependence between $x$=3$\%$ and $x$=5$\%$, consistent with transport measurements, and the onset of a (granular) metal phase (Fig.\ref{cluster}). For doping levels above 5$\%$, signal wipeout is present at all temperatures, but a change in spectral shape and both $T_1$ and $T_2$ relaxation rates occurs at low temperatures. Below 100~K, spectral lines for samples with $x$=7$\%$ and 8$\%$ broaden and wings appear at line edges, which show a qualitatively different temperature and frequency evolution than those measured at line center. Even though NMR can not provide direct insight into the geometry and spatial configuration of charges, our measurements in this region of the phase diagram show evidence of distinct electronic environments, and shed light on the microscopic evolution of metallicity and short-range spin correlations in lightly doped cuprates.

The paper is organized as follows: In Section II, we describe the results of our Cu NMR measurements, including spectra, spin-lattice and spin-spin relaxation times; in Section III, we discuss the NMR results and their implications, and Section IV provides a summary.

%  \cite{Fujita2002, Wakimoto2000, Braicovich2010, Niedermayer1998}. 

\section{Experiment and results}
\label{NMR results}

For our Cu NMR measurements, we used high-quality oriented single crystals with hole doping levels $x$ = 2$\%$, 3$\%$, 5$\%$, 7$\%$, 8$\%$ and 19$\%$. All crystals were cut from boules grown by the traveling floating zone method, oriented with x-ray Laue diffraction and subsequently annealed using standard procedures. Superconducting samples were characterized using SQUID magnetometry, and their T$_c$ values agree well  with prior work. \cite{Wen2019, Badoux2016}
Our measurement focused mostly on the isotope $^{63}$Cu, due to its higher abundance compared to $^{65}$Cu. Both copper nuclei have spin $3/2$ and non-zero quadrupolar moments, which implies that the NMR spectrum for each isotope consists of a central line and two quadrupolar satellites. Since we were primarily interested in spin fluctuations, we did not study quadrupolar effects, and all measurements were conducted on the central NMR line, which corresponds to the $-\frac{1}{2}\rightarrow \frac{1}{2}$ nuclear spin level transition. The experiments were performed using Tecmag broadband spectrometers and a high-homogeneity 12~T superconducting magnet. The crystals were oriented with the external magnetic field, $H_0$, parallel to the crystallographic $ab$ planes, in order to minimize signal wipeout effects (see below). We employed standard spin-echo sequences with echo times as short as 2~$\mu$s. This required fast-recovery signal preamplifiers, significant damping in the resonant circuit that contains the NMR coil, and phase cycling to minimize the effects of resonant circuit ringing after the NMR pulses  \cite{Pelc2017}. Spin-lattice relaxation times were determined using standard saturation recovery pulse sequences, and the relaxation was assumed to be due to purely magnetic interactions, similar to previous work  \cite{Pelc2017, Imai2017}. A magnetic relaxation mechanism has been established in LSCO-15$\%$ by Walsted et al. through measurements of $T_1$ at both the $^{63}$Cu and $^{65}$Cu spectral lines  \cite{Walstedt1995}. Also, measurements at high temperatures have shown Curie-Weiss behavior for underdoped LSCO \cite{Fujiyama1997}.

\subsection{$^{63}$Cu NMR spectra}
%\subsubsection{Room temperature measurements - line shape and position}

\begin{figure*}
\includegraphics[trim={4.5cm 0cm 5.5cm 0cm},clip, width=1\linewidth]{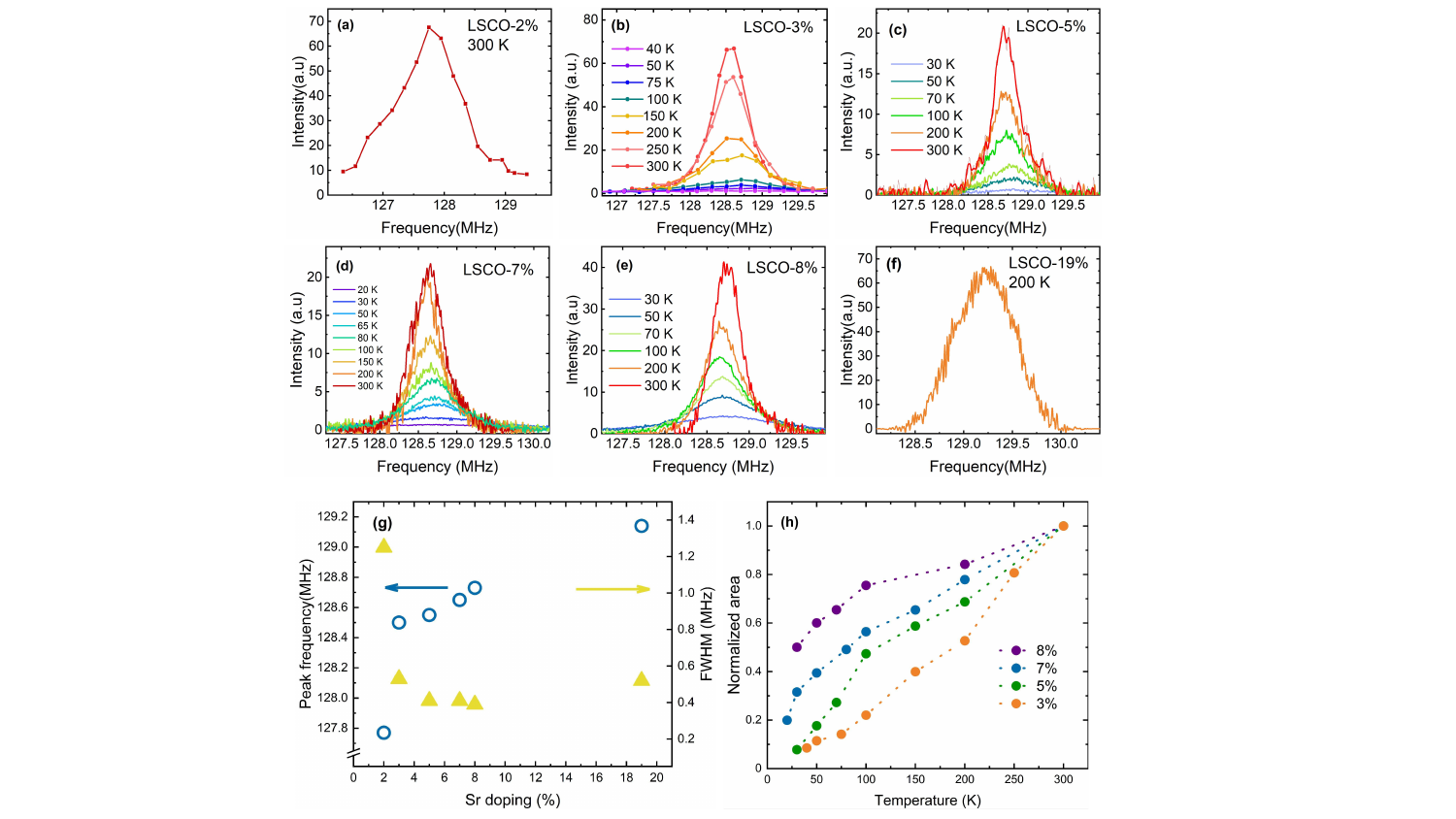}
\caption[Spectral lines at different temperatures]{Central NMR transition lines at different temperatures for samples with doping levels (a) $x$ = 2$\%$, (b) $x$ = 3$\%$, (c) $x$ = 5$\%$, (d) $x$ = 7$\%$, (e) $x$= 8$\%$ and (f) $x$ = 19$\%$. Similar signal wipeout has been observed for all samples that have been measured at several temperatures. While there is line broadening for $x$ = 7$\%$  and $x$ = 8$\%$ , where characteristic "wings" appear, no such effect is observed for $x$ = 3$\%$ and $x$ = 5$\%$. (g) Peak positions (circles) and FWHM values (triangles) at 300~K for LSCO crystals with various doping levels. The peak frequency increases with doping, and between $x$ = 3$\%$ and 8$\%$ the relative values are determined by the second-order quadrupolar contribution, unlike the sudden change between $x$ = 2$\%$ and 3$\%$ that cannot be accounted for by the quadrupolar shift alone. The peak frequency measured for LSCO-5$\%$ has been scaled by a factor 0.998, as the magnetic field for that experiment was set to 11.22~T. % A linear doping dependence of $\nu_Q$ has been assumed, and the values extrapolated from measurements on dopings higher than  10$\%$  \cite{Ohsugi1994}. 
FWHM values mirror the center frequency result: LSCO-2$\%$ has a significantly wider central line than the rest of the samples, whose widths do not vary greatly with doping. (h) Integrated raw spectral intensities normalized to the values at 300~K (even though the slope indicates that wipeout onset temperatures are higher for all samples). The integration spanned the whole measured range, and no $T_2$ corrections have been made. Signal wipeout is consistently stronger at lower doping, and for samples with $x >$ 3$\%$, at lower temperatures.}
\label{wipeout} \label{wipeout-spectra} 
\end{figure*}

The temperature dependence of the central NMR transition line was measured down to 20-30~K for all underdoped samples except LSCO-${2\%}$, for which only an ambient-temperature measurement could be performed due to strong signal wipeout (Fig. \ref{wipeout} (a)-(e)). For samples with doping levels between 3$\%$ and 8$\%$, the frequency of the central line is  128.5-128.8~MHz in an applied field of 11.2~T, and increases with doping, as seen in Fig. \ref{wipeout} (g). This shift is likely dominated by second-order quadrupolar contributions that are known to be present with the external field oriented perpendicular to the crystallographic $c$-axis. Because of our setup geometry, in which the external field is oriented to lie in the Cu-O planes, this shift is dominated by the second-order quadrupolar contribution, which can be calculated from the known expression \cite{Walstedt2018}:

\begin{equation}
\label{qudropolar shift}
\nu^{(2)}_{\pm \frac{1}{2}}=\nu_L-\frac{\nu_Q^2}{16\nu_L}\left[I(I+1)-\frac{3}{4}\right]\sin^2\theta_c(9\cos^2\theta_c-1).
\end{equation} 

where $\nu_L$ and $\nu_Q$ are the Larmor and quadrupolar frequencies, respectively, $I = 3/2$ is the $^{63}$Cu nuclear spin, and $\theta_c$ is the angle between the magnetic field and the crystallographic $c$-axis. For LSCO-7$\%$ with in-plane field, i.e., $\theta_c$=90$^\circ$, the frequency of the peak is 128.7~MHz. Using $\nu_Q$=34~MHz  \cite{Imai1993, Ohsugi1994} and expression (\ref{qudropolar shift}), the quadrupolar shift can be estimated to be 1.7~MHz. The total shift is 2.2~MHz, which gives a Knight shift estimation of K$_{90}$=0.5~MHz, (K$_{90}$=0.4$\%$). Comparing these two values, we can see that in our sample orientation the quadrupolar term is significantly larger than the Knight shift.

The doping-induced increase of the peak frequency is consistent with a linear change of the quadrupolar frequency which was previously established at higher doping levels ($x>$10$\%$) \cite{Ohsugi1994}. On the other hand, in LSCO-2$\%$ the line is centered at 127.8~MHz -- a large shift that suggests an abrupt change in either quadrupolar or magnetic shifts (or both). Similarly, the width (full width at half maximum, FWHM) of the spectral lines is $\sim$0.5~MHz at all doping levels except LSCO-2$\%$, for which it is 1.3~MHz (Fig \ref{wipeout} (g)). Since the LSCO-2$\%$ crystal was grown and characterized in the same manner as other samples, the broadening is unlikely to be a result of extrinsic imperfections, but an intrinsic property of the system. The wide spectral lines measured in  LSCO-2$\%$ are in stark contrast with those measured in the undoped parent compound, where the signal is very narrow  \cite{Imai1993}.
We also note that we did not observe any signature of the additional NMR line associated with Cu nuclei close to Sr dopants, known as the B line, in any of the studied samples. This agrees with prior NQR experiments which show that the B line is essentially undetectable below $x$=10$\%$.  \cite{Ohsugi1994}

%\begin{figure}[t!]

%\includegraphics[width=\linewidth]{slike/FpeakpositionsFWHM.eps}
%\caption[Peak positions and FWHM at 300~K for various doping]{Peak positions (circles) and FWHM values (triangles) at 300~K for LSCO crystals with various doping levels. The peak frequency increases with doping, and between $x$ = 3$\%$ and 8$\%$ the relative values are determined by the second-order quadrupolar contribution, unlike the sudden change between $x$ = 2$\%$ and 3$\%$ that cannot be accounted for by the quadrupolar shift alone. The peak frequency measured for LSCO-5$\%$ has been scaled by a factor 0.998, as the magnetic field for that experiment was set to 11.22~T. % A linear doping dependence of $\nu_Q$ has been assumed, and the values extrapolated from measurements on dopings higher than  10$\%$  \cite{Ohsugi1994}. 
%FWHM values mirror the center frequency result: LSCO-2$\%$ has a significantly wider central line than the rest of the samples, whose widths do not significantly vary with doping.}
%\label{peak position}
%\end{figure}

%\subsubsection{Temperature dependence - signal wipeout} 

Upon cooling, signal wipeout is observed in all samples, as shown in Fig. \ref{wipeout-spectra}. For $x$ = 7$\%$ and especially $x$ = 8$\%$, the lines broaden at temperatures below 100~K, and the lineshape is no longer Gaussian, as characteristic wings appear at both the high- and low-frequency sides. This additional signal component has been observed in samples with $x$ = 6$\%$ and  $x$ = 11.5$\%$ in previous work  \cite{Julien1999, Imai2017}, and will be discussed in more detail in Section III.

Because of variations in the experimental setup, coil characteristics and sample sizes, the raw signal intensities obtained for different samples are not directly comparable. However, our main conclusions do not depend on this. In order to determine the wipeout characteristics, all spectral lines were integrated in the entire measurement range, corrected for the Boltzmann factor, and the areas normalized to ambient temperature values. This heuristic procedure has its limitations, since the wipeout clearly onsets at higher temperatures, especially for samples with the lowest doping levels. However, the normalization is still useful to show semi-quantitatively that wipeout is stronger for smaller $x$, with a possible change in slope at lower temperatures (Fig. \ref{wipeout} (h)). At this point no $T_2$ corrections were made, and only raw data is shown, which implies that the true signal intensity is larger. The effects of $T_2$ are discussed in Section III.

\subsection{Nuclear relaxation times}

Previous work on La-based cuprates has shown that spin-spin relaxation times $T_2$ are relevant for the wipeout phenomenon -- short $T_2$ values directly impact the spin echo intensities that can be detected with finite echo times -- while spin-lattice relaxation times $T_1$ tend to be significantly longer  \cite{Pelc2017,Imai2017, Imai2021}. To investigate the origin of signal wipeout, we have conducted systematic $T_2$ measurements across the spectral lines and in a wide range of temperatures for all underdoped samples, using a standard spin echo sequence. In order to pinpoint the origin of the dependence of $T_2$ on temperature and frequency, as well as determine the Redfield component of $T_2$, spin-lattice relaxation times were also measured, for two samples: LSCO-7$\%$ and LSCO-8$\%$. %All values described below have been obtained by a single exponential fit.
%\begin{equation}
%f(t)=M_0 \text{e}^{-t/T_2}.
%\end{equation}
\begin{figure*}[t!]
\includegraphics[trim={1cm 2cm 1cm 2cm},clip, width=0.9\linewidth]{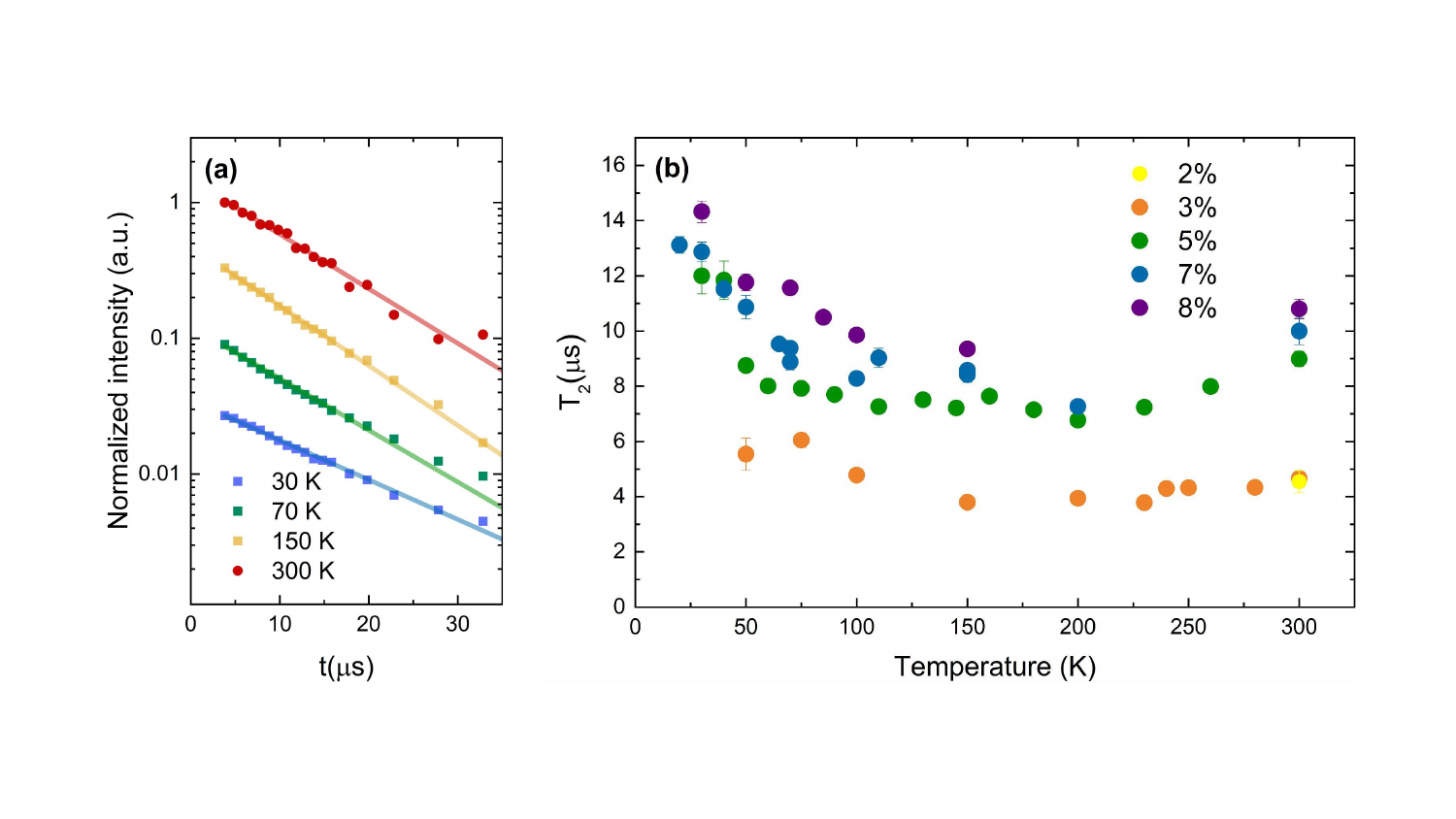}

 \caption[]{\label{T2_fitovi} \label{T2_temp_svi} (a) Relaxation curves measured at the line center for LSCO-8$\%$ at different temperatures. All curves are purely exponential, as is expected for this sample orientation \cite{Imai2017}. (b) Temperature dependences of spin-spin relaxation times obtained from single-exponential fits measured at the peak frequency. $T_2$ shortens as doping decreases, and is rather small for the lowest doping levels. In the temperature range from 300~K to 150~K $T_2$ shortens for all samples, but upon further cooling the relaxation slows down.}
\end{figure*}
In Figure \ref{T2_fitovi} (a), relaxation curves  measured at different temperatures are shown for LSCO-8$\%$. The temperature dependence of spin-spin relaxation times for underdoped samples is shown in Figure \ref{T2_temp_svi} (b). In general, $T_2$ decreases with decreasing doping level, and this effect is especially pronounced between $x$ = 5$\%$ and 3$\%$. Furthermore, $T_2$ shortens upon cooling from 300~K to 200~K, and exhibits an upturn at lower temperatures, which is more pronounced for samples with higher doping levels.
\begin{figure}[t!]

\end{figure} 
\begin{figure}[t!]
\centering
\hspace*{0 cm}
\includegraphics[width=0.8\linewidth]{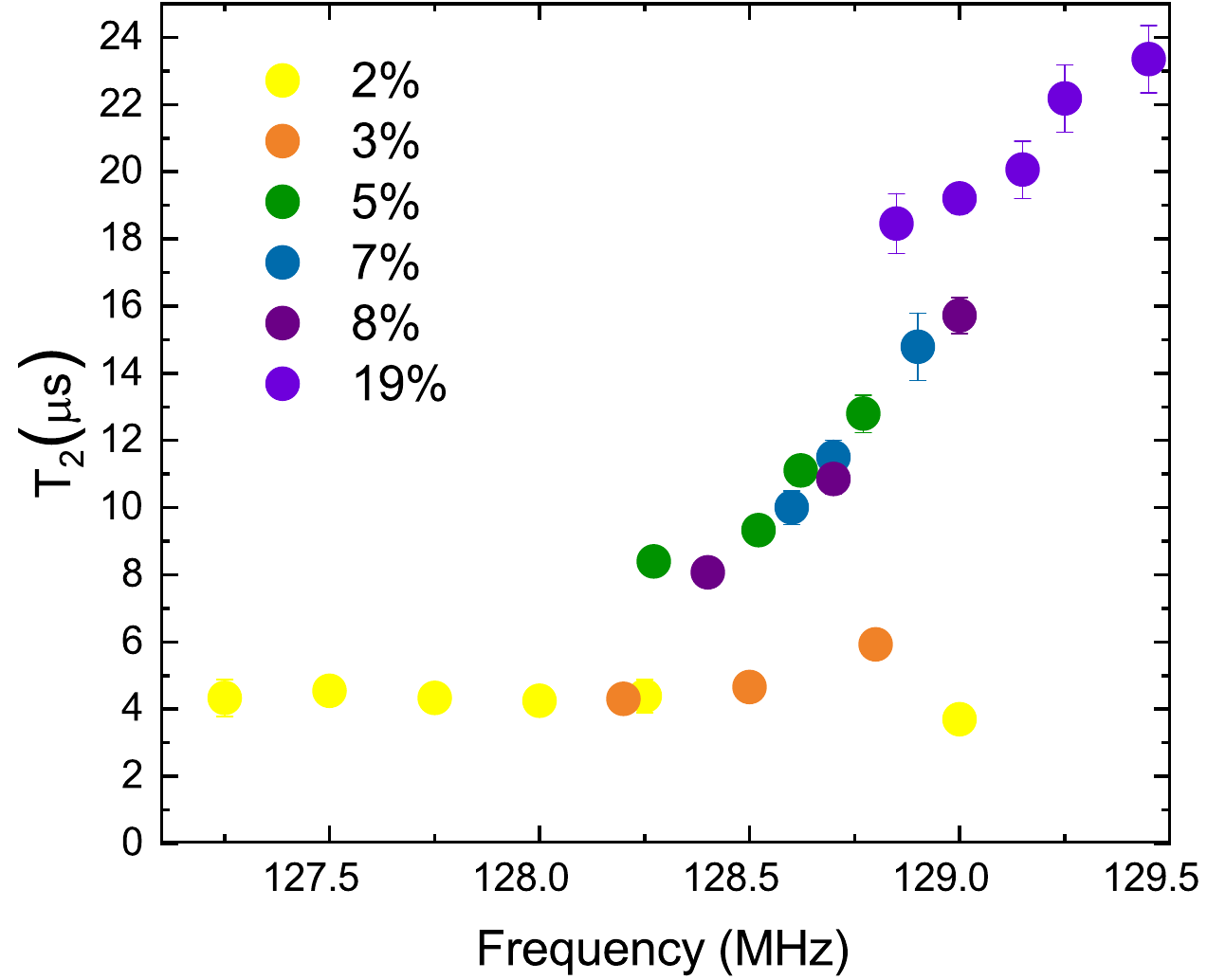}
\caption[T$_2$ across the spectral lines for samples with doping levels 2$\%$ to 19$\%$, at 300~K.]{\label{t2300k} Spin-spin relaxation rates obtained from single-exponential fits at 300~K and different frequencies. Three groups are clearly discerned:  strongly underdoped samples relax faster, with less variation across the spectral line. Samples with $x$ = 5$\%$ , 7$\%$  and 8$\%$ show similar behavior, and $x$=19$\%$ relaxes significantly slower. The observed trends are not a simple consequence of doping heterogeneity, since signals at the same frequencies relax differently for different groups.}
\label{T2tempfrek}
\end{figure} 
Previous NQR measurements have shown that $T_2$ in LSCO varies across the spectral line, which has been attributed to doping heterogeneity,  \cite{Hunt1999,Singer2002}. In Figure \ref{t2300k}, we show $T_2$ measured at several frequencies around the central peak, at ambient temperature. As expected, the relaxation times generally increase with doping and frequency across the spectral line.  We find, however, that doping heterogeneity effects cannot entirely explain the behavior of $T_2$ in strongly underdoped samples. For samples with $x$ = 5$\%$, 7$\%$ and 8$\%$, the $T_2$ values overlap for different samples at the same frequency, which is consistent with doping heterogeneity, and the scaling even roughly extends to the overdoped LSCO-$19\%$ sample. On the other hand, this is clearly not valid for the strongly underdoped, non-superconducting samples with $x$ = 2$\%$ and 3$\%$. Although we expect long-range antiferromagnetic order to be nearly completely suppressed in LSCO-2$\%$, the $T_2$ values for this composition are significantly shorter than those of higher-doped samples at corresponding frequencies. Furthermore, there is a qualitative change in the variation of $T_2$ across the spectral line: unlike in samples with higher doping, $T_2$ in LSCO-2$\%$ remains constant at all measured frequencies. LSCO-3$\%$, on the other hand, shows similarities with both lower- and higher-doped samples. While the relaxation rate seems to roughly follow the general frequency trend seen in samples with higher doping, the $T_2$ values are significantly shorter overall, closer to LSCO-2$\%$. This cannot be a simple consequence of doping heterogeneity, since $T_2$ in LSCO-3$\%$ is up to a factor of two shorter than the $T_2$ values measured in LSCO-8$\%$ at the same frequency.

\begin{figure}[t!]
\centering
\includegraphics[trim={8.5cm 0.2cm 8.5cm 0.2cm}, clip, width=0.81\linewidth ]{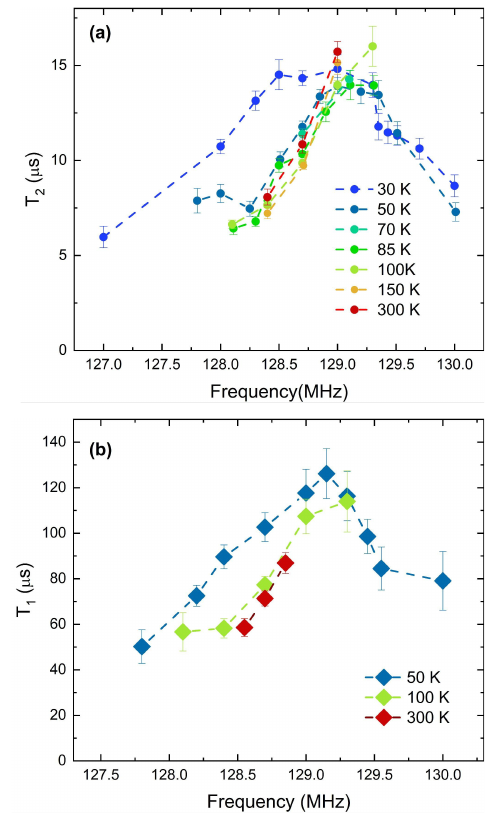}
   
\caption {\label{T1frek}(a) Spin-spin relaxation rates obtained from single exponential fits to the echo decay, for LSCO-8$\%$ at different frequencies in the range from 127~MHz-130~MHz at some of the measured temperatures. A characteristic pattern can be observed: across the spectral line the relaxation is linear at high temperatures, similar to known NQR results. Upon cooling, $T_2$ exhibits a downturn at higher frequencies, and the relaxation at the peak slows down. At low temperatures, both line edges relax significantly faster than the central part, resulting in a peak-like shape. As the spectral line broadened upon cooling, the frequency range of $T_2$ measurements increased. (b) T$_1$ relaxation times for  LSCO-8$\%$ at various temperatures measured at different frequencies across the spectral line. T$_1$ and  T$_2$ temperature and frequency trends are similar, as fast relaxing zones appear at both line edges at low temperatures.}
\label{t2tempfrek}
\end{figure}

\begin{figure}[t!]
\centering
\hspace*{0 cm}
\includegraphics[trim={0cm 0cm 0cm -0.2cm}, clip, width=0.74\linewidth]{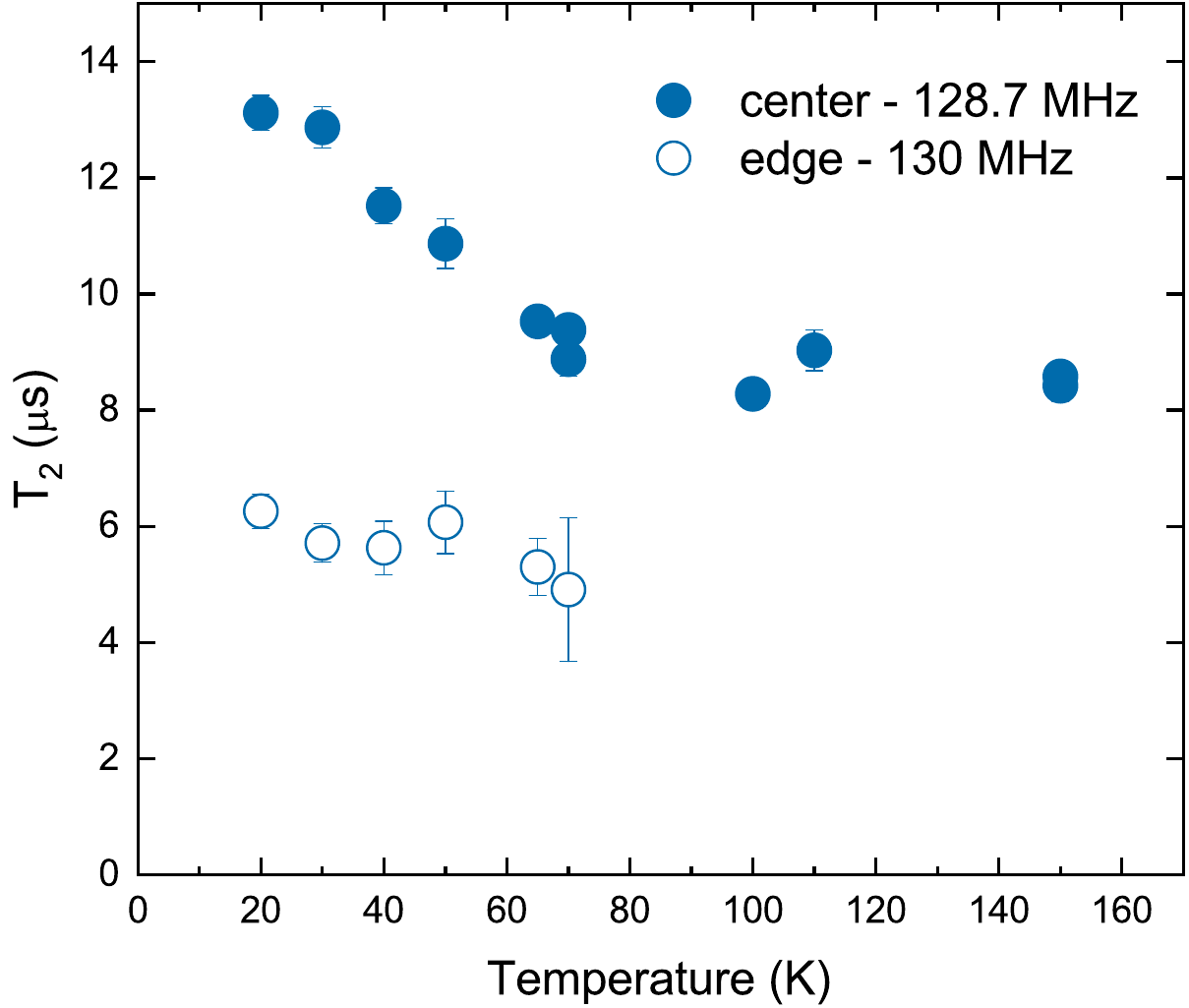}
\caption[LSCO-7$\%$ relaxation rates]{ Spin-spin relaxations for LSCO-7$\%$ measured at the line peak and edge. The temperature dependences are qualitatively different, which implies the presence of distinct local electronic environments. }
\label{t2centeredge}
\end{figure}

Because of this nontrivial behavior, we also measured changes of $T_2$ with temperature at various frequencies across the spectral lines. Upon cooling, two effects can be discerned. First, the relaxation time measured at the line center increases below 100~K. Second, the longer relaxation times at the high-frequency end of the spectral line become shorter. As shown in Fig. \ref{t2tempfrek} (a) for LSCO-8$\%$, the frequency dependence of $T_2$ exhibits a gradual downturn upon cooling, crosses over from roughly linear to S-shaped, and finally develops a peak at the lowest temperatures, where both the high- and low-frequency edges relax significantly faster than the line center. Similar behavior is observed in LSCO-7$\%$ at two frequencies, 127.7~MHz and 130~MHz, corresponding to the line peak and wing, respectively, shown in Fig.\ref{t2centeredge}. The temperature dependence of the peak and wing relaxation rates are qualitatively different, which suggests that electronically distinct environments emerge.


%\begin{figure}[t!]
%\centering
%\begin{subfigure}{\linewidth}
% 	\centering
% 	\hspace*{0 cm}
%	\includegraphics[trim={1.5cm 1.5cm 1.5cm 1.5cm}, clip,  width=0.45\linewidth]{slike/T230K}
%	\label{t2spectrum7}
%\end{subfigure}
%\centering
%\hspace*{0 cm}
%\includegraphics[trim={1.5cm 1.5cm 1.5cm 1.5cm},clip, width=0.55\linewidth, ]{slike/T250K}
%\caption[Relaxation rates at 50~K for LSCO samples doped $x$=5$\%$, 7$\%$, 8$\%$]{Relaxation rates at 50~K for LSCO samples doped $x$=5$\%$, 7$\%$ and 8$\%$. The same frequency dependence can be seen.}
%\label{t2frek5}
%\end{figure}

\begin{comment}
\begin{figure}[t!]
\centering
\includegraphics[width=1\linewidth]{slike/t1_lsco8}
\caption[T1 relaxation curves for LSCO-8$\%$ at different temperatures]{\label{T1 curves} Representative T$_1$ recovery curves for LSCO-8$\%$ at different temperatures. Raw data is shown, so final intensities are not corrected for temperature and acquisition details.}
\end{figure}
\end{comment}
\begin{figure*}[t!]
\centering
\hspace*{0 cm}
\includegraphics[trim={0.8cm 2cm 0.9cm 2cm},clip, width=0.9\linewidth]{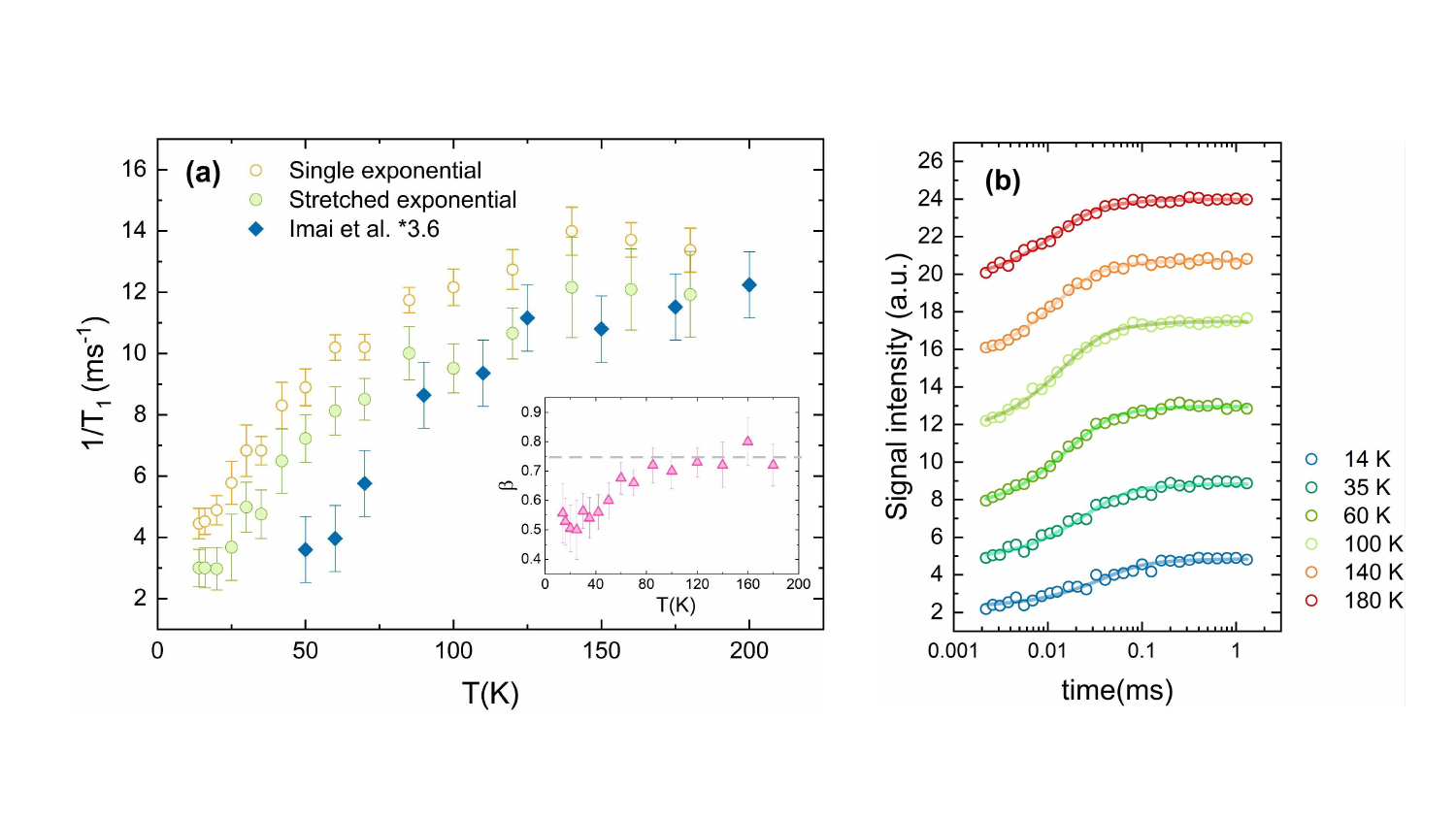}

\caption[LSCO-8$\%$ relaxation rates]{\label{T1_fitovi} (a) Temperature dependence of 1/T$_1$ for LSCO-8$\%$ at the center of the spectral line. The values were obtained from single exponential curves (orange symbols) and from stretched exponential curves (green symbols). Both methods yield relaxation rates slightly higher than those measured by Imai \cite{Imai1990} (blue symbols), when the anisotropy factor of 3.6 is included. The stretching coefficient $\beta$ is shown in the inset. (b) Spin-lattice magnetization relaxation curves measured at line center for LSCO-8$\%$ at different temperatures. The fitted curves are single exponential. }
\end{figure*} 
Spin-lattice relaxation measurements were performed mostly for LSCO-8$\%$, and to a lesser extent for LSCO-7$\%$. Qualitatively, $T_1$ shows similar frequency and temperature dependences as $T_2$, with $T_1$ roughly an order of magnitude longer. The temperature dependence of 1/$T_1$ at the line center, shown for LSCO-8$\%$ in Fig. \ref{T1_fitovi} (a), is a smooth curve that shows a gradual decrease in cooling, similar to spin-lattice dynamics seen in samples with higher doping  \cite{Gorkov2004, Baek2017, Smith2006}. Generally, $T_1$ is shorter than for higher doping levels  \cite{Avramovska2020, Jurkutat2015}. The exact values of $T_1$ depend on the fit function: we have determined the values by fitting both single exponential and stretched exponential functions, with the latter generally yielding lower mean relaxation rate values. The stretching coefficient $\beta$ at higher temperatures is about 0.75, and it starts to decrease rapidly below 80~K, about the same temperature below which wings can be discerned in the spectra. This is consistent with the development of strong electronic inhomogeneity. The mean values can be compared to results by Imai et al. \cite{Imai1990}, after correcting for an anisotropy factor of 3.6. At higher temperatures the values are relatively close, but below 100~K our relaxation rates are significantly faster. This could be due to the fact that the short echo times allow us to pick up faster components than in previous measurements. Since the $T_1$ values obtained from stretched and single exponential fits are fairly close, in order to minimize the number of fitting parameters we use simple exponential fits for further analysis below.The relaxation curves, and single exponential fits, are shown in Figure \ref{T1_fitovi} (b).In addition, measurements across the spectral line show that $T_1$ displays a similar frequency dependence as $T_2$ (Fig. \ref{T1frek} (b)). At higher temperatures, $T_1$ increases with frequency, consistent with previous NQR measurements in this doping range, which were attributed to an intrinsically inhomogeneous electronic state \cite{Fujiyama1997}. Upon cooling we observe the same crossover from a roughly linear to a peak-shaped frequency dependence as for $T_2$.

\section{Discussion} 

\subsection{Correction of spectral intensity for relaxation rates}

\begin{figure*}[t!]
    \centering
        \includegraphics[trim={0cm 3.5cm 0cm 3.5cm},clip, width=1\linewidth]{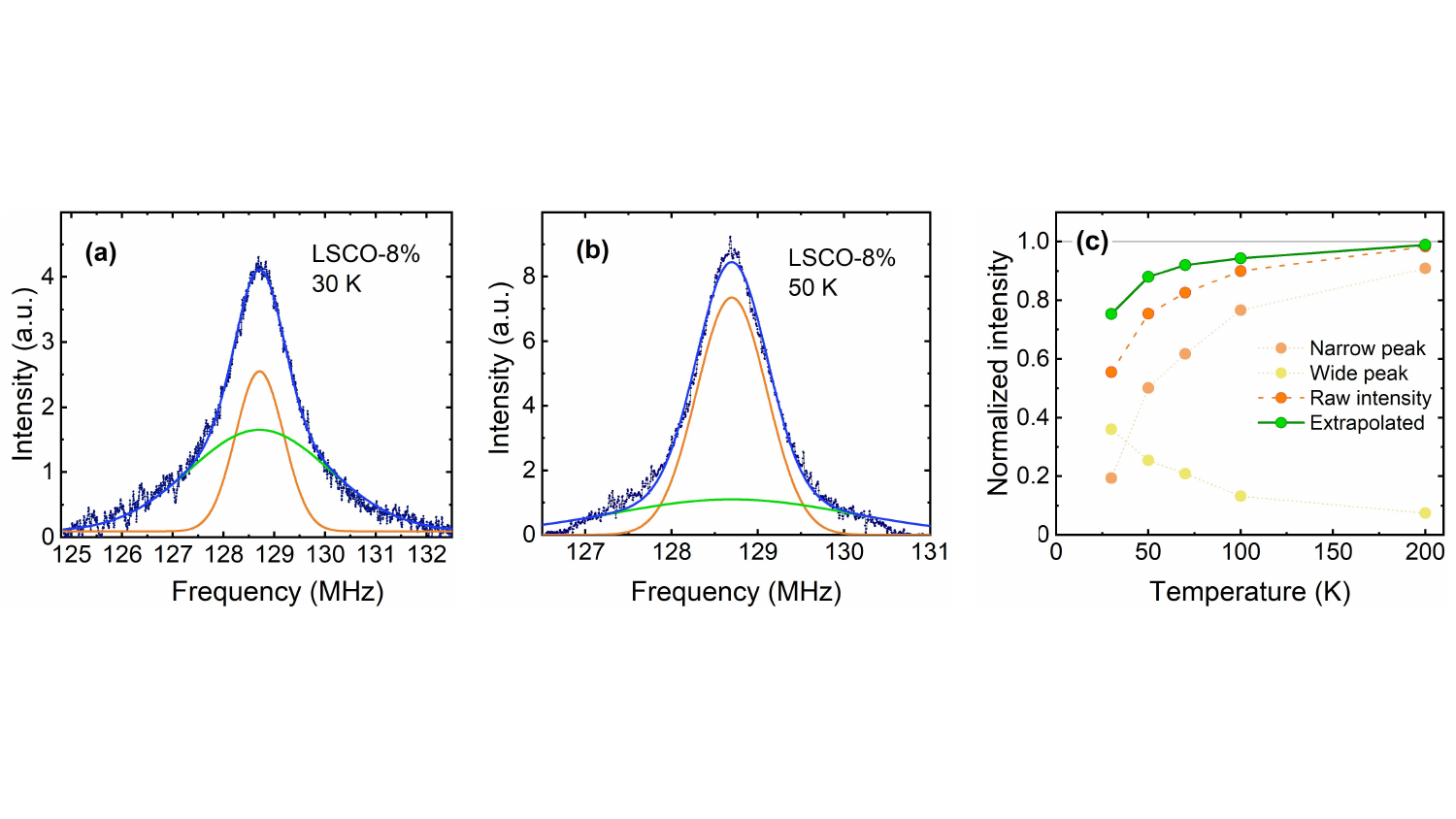}
    \caption[Two line model] {Two Gaussian peaks, fitted to the spectra of LSCO-8$\%$ at (a) 30~K and (b) 50~K. The width of the wider peak was kept constant, and taken from the fit at the lowest temperature. The width of the narrow peak increased with cooling.\label{gausspeak} (c) Temperature dependence of the areas of two Gaussian peaks obtained from the spectra for LSCO-8$\%$. For the narrow peak, $T_2$ measured at the line center at 200~K was used, wehereas for the wide peak we used $T_2$ at the line edge and at 30~K. \label{gaussextrapolacija} }
\end{figure*}

The non-Gaussian broadening of spectral lines, together with qualitatively different $T_2$ trends at the center and edges of the spectral lines, indicate the presence of electronically distinct environments in LSCO-7$\%$ and 8$\%$  at low temperatures. To explore this further, we use a minimal two-component model: we fit two Gaussian curves with the same peak positions and different widths (roughly $<$1~MHz and 2.8~MHz) to the spectra, similar to the approach by Julien \textit{et al.} [\citenum{Julien1999}]. Representative fits for LSCO-8$\%$ are shown in Fig. \ref{gausspeak} (a) and (b). At temperatures above 100~K, the weight of the broad component is less than 10$\%$ of the weight of the narrow component, while at the lowest measured temperature of 30~K, the broad peak weight becomes about 65$\%$.

These fits can be used to estimate the number of nuclei that contribute to the signal at each temperature, by correcting for the varying spin-spin relaxation rates. As the relaxation rates at both the high- and low-frequency edges of the spectral line are significantly faster than in the central part, it is plausible to attribute a single, fast relaxation time to the entire broad Gaussian component. In order to extrapolate the spin echo intensity to $t=0$, we then assign the $T_2$ values measured at the wings to the broad/fast component, and the $T_2$ values measured at the line center to the narrow/slow component. This is not entirely correct, since the fast-relaxing phase contributes to the relaxation rate measured at the center, implying that the actual relaxation rate of the slow phase is overestimated. However, the narrow component is sufficiently stronger at the line center that this correction does not significantly change the results.

In order to minimize the number of free parameters in the fits, we keep the width of the broad peak fixed to the value determined at the lowest measured temperature, 2.8~MHz, while the narrow peak width was unconstrained. The areas under the curves were extrapolated to time $t=0$ assuming simple exponential decay of the spin echo intensity, and the results are shown in Fig. \ref{gaussextrapolacija} (c). As can be seen, the wipeout fraction of the extrapolated intensity is smaller than that of the raw measurements with finite echo times, but wipeout is still present, especially at lower temperatures.
\begin{figure}[t!]
\centering
\includegraphics[trim={0 0 0 0},clip, width=0.8\linewidth]{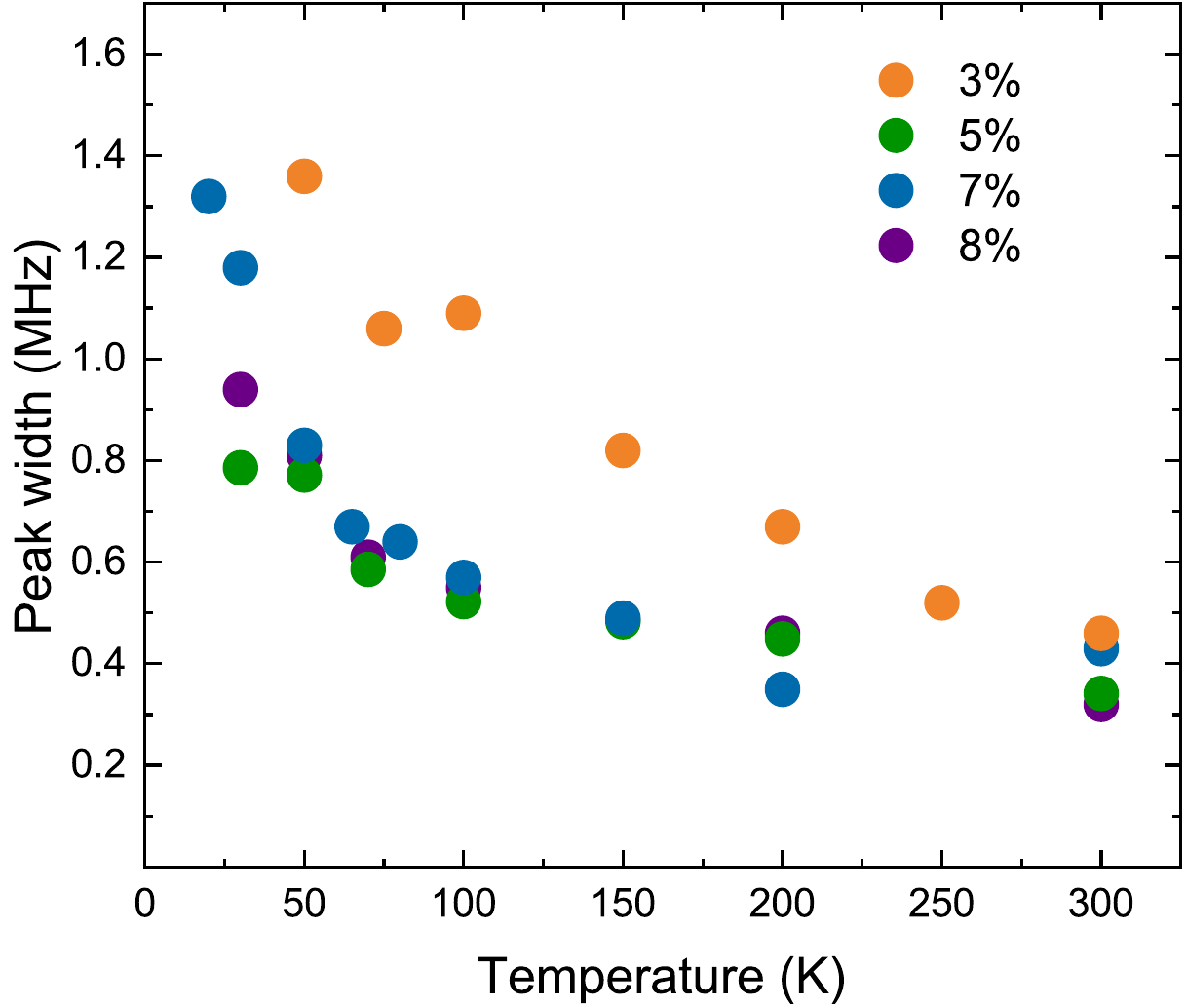}
\caption[Two line model - wipeout] {The width of the narrow gaussian peak fitted to the spectral lines. For LSCO-5$\%$ and LSCO-3$\%$ only one peak has been fitted, as no clear wing signal has been observed. \label{peakwidth}}

\end{figure}
These results are similar to the spectral intensity extrapolations performed by Imai \textit{et al.} (ref. \citenum{Imai2017}) for LSCO $x = 11.5$\%, which show some recovery of the signal intensity at higher temperatures where charge order and fluctuations dominate, but still find wipeout for all echo times below 100~K.

We can therefore conclude that wipeout in underdoped LSCO is likely not just due to fast spin-spin relaxation, since the extrapolation does not account for the entire missing signal. This is in contrast to e.g. La$_{2-x}$Ba$_x$CuO$_4$ around $x$ = 12.5$\%$, where short $T_2$ values have been shown to explain the apparent loss of signal \cite{Pelc2017, Singer2020, Imai2021}. It is possible that a component with even faster relaxation rates is present in underdoped LSCO, which would be completely unobservable in our experiments, or that the spectral lines are significantly broader than what we could detect. There is no simple way to verify either of these scenarios, as our detection times are limited by the electronics, and even with more detailed measurements and longer acquisition times it would be difficult to discern such small and broad signals. 

Although it does not completely account for the wipeout effect, it is interesting to consider an effective two-component picture, as sketched in Fig. \ref{cluster}. A comparison of the narrow peak widths determined from the fits is shown in Figure \ref{peakwidth}. The widths overlap for $x$ = 5$\%$, 7$\%$, and 8$\%$, even at low temperatures, where the narrow peak weight is relatively small. The similarity of the widths indicates that the underlying electronic environments are similar as well. In contrast, in LSCO-3$\%$ the line is significantly broader across the entire measured temperature range. This suggests a qualitative change between 3\% and 5\%, consistent with the behavior of $T_2$, and the picture presented in Fig. \ref{cluster}.

\subsection{The origin of $T_{2}$}
\label{t2analysis}
In general, there are two contributing factors to the spin-spin relaxation rate $T_2$: the intrinsic spin-spin relaxation, and the Redfield correction $T_{2R}$ from spin-lattice relaxation  \cite{Imai2017}. With the external field parallel to the crystallographic $c$-axis, the intrinsic component $T_{2G}$ has a Gaussian form, and can be determined from a direct fit to the decay curve, either by fixing the Redfield correction from measured values of $T_1$, or by treating it as a free parameter  \cite{Imai2017, Singer1999}. 
For in-plane magnetic field, where spin-flip processes are important for decoherence,  Pennington-Slichter theory predicts an exponential intrinsic relaxation $T_{2E}$, which results in single-exponential relaxation curves and makes the distinction of terms less straightforward  \cite{Pennington1989}. The total exponential decay function can be written as: 
\begin{equation}
\label{T2r}
f(t)=M_0\text{e}^{-t/T_{2R}}\text{e}^{-t/T_{2E}}.
\end{equation} 
Since the decays associated with $T_{2E}$ and $T_{2R}$ have the same functional form, they cannot be distinguished from numerical fits alone, and in order to determine $T_{2E}$, the Redfield correction must be estimated independently. For an in-plane geometry, $T_{2R}$ is connected to both in-plane $T_{1ab}$ and out-of-plane $T_{1c}$ through the following expression \cite{Kramer2002}:

\begin{equation}
\label{T2rKramer}
\frac{1}{T_{2R}}=\frac{1}{2}\frac{1}{T_{1c}}+\frac{7}{2}\frac{1}{T_{1ab}}.
\end{equation}

In our case, we can use measured and published values of $T_{1ab}$ and $T_{1c}$ to estimate $T_{2R}$ to be $\sim$ 20~$\mu$s for LSCO-8$\%$. The remaining contribution to the relaxation then results from indirect spin-spin interactions. For LSCO-15$\%$, the characteristic relaxation time that corresponds to this mechanism has been estimated theoretically and experimentally by Walstedt to be $\sim$ 50 $\mu$s  \cite{Walstedt1996}, with the magnetic field along the $c$-axis. If we take into account the anisotropy of the system, the spin flip rate changes as (1+3 $\sin \theta_c$), where $\theta_c$ is the applied field angle in regard to the $c$ axis  \cite{Walstedt2018}. Therefore, the spin-spin relaxation time is significantly shorter for an in-plane field orientation,  $\sim$ 12 $\mu$s. When we combine this estimate with the Redfield correction for LSCO-8$\%$, the total predicted relaxation time becomes $T_{2est}\sim$ 8~$\mu$s. Our experimental result for $T_2$ in LSCO-8$\%$ above 100 K, $T_2 \sim 10~\mu$s, agrees well with the predicted value.

\begin{figure*}[t!]

\includegraphics[trim={0cm 2cm 0cm 1cm},clip, width=0.9\linewidth]{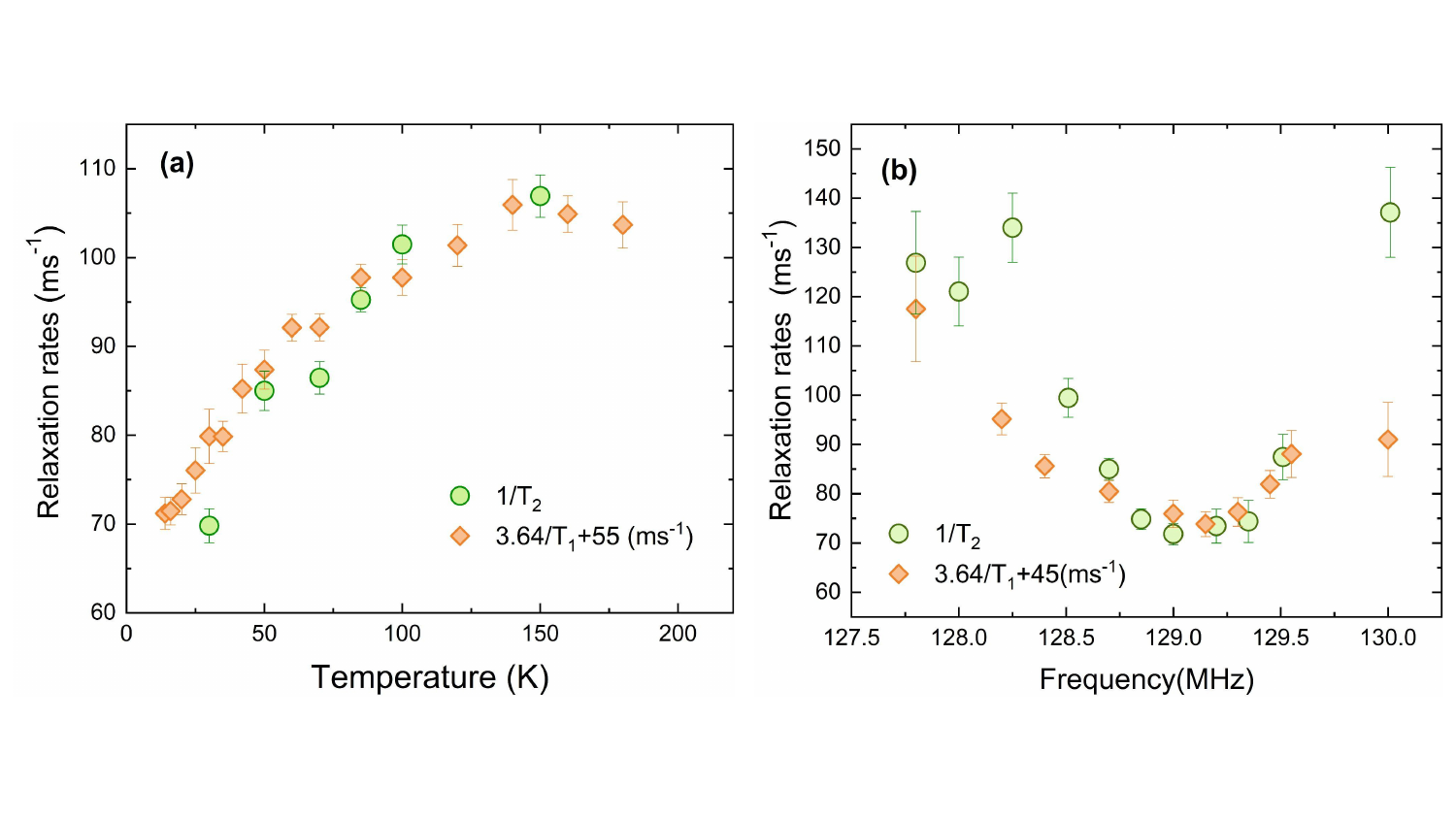}
\caption[Relaxation rates for LSCO-8$\%$ measured at 50~K, offset]{(a) 1/T$_1$ (diamonds) and 1/T$_2$ (circles) relaxation rates for LSCO-8$\%$ in dependence on temperature. (b) 1/T$_1$  and 1/T$_2$  for LSCO-8$\%$ at 50~K in dependence on frequency. $1/T_1$ values are scaled and shifted to match $1/T_2$ as described in the text.\label{t1t2frek} \label{t1t2temp} }

\end{figure*}

In order to further elucidate the relaxation mechanisms, we can perform a detailed comparison of the extensive $T_1$ and $T_2$ data for LSCO-8$\%$ and LSCO-7$\%$, obtained in a wide range of temperatures and across the spectral line. From expressions (\ref{T2r}) and  (\ref{T2rKramer}) and using the known $T_1$ anisotropy factor of $3.7$ (ref.  \citenum{Ohsugi1994}), we can derive a relation between the two observables $T_{2ab}$ and $T_{1ab}$:
\begin{equation}
\frac{1}{T_{2ab}}=\frac{1}{T_{2E}}+ \frac{3.63}{T_{1ab}}.
\end{equation}
The anisotropy factor is derived from our measurements at room temperature and $T_1$ results by Ohsugi 
\textit{et al.}, and agrees with values obtained for LSCO-11.5$\%$ and undoped 
 La$_2$CuO$_4$ \cite{Imai2017,Imai1993, Imai1993b, Imai1993-2}. The simplest assumption we can make is that $T_{2E}$ is independent of temperature, which would imply that the temperature dependence of $T_{2ab}$ is solely due to $T_1$, which enters through the Redfield term. The measured relaxation rates for LSCO-8$\%$ indeed collapse on a single curve when offset by a constant factor, as shown in Fig. \ref{t1t2temp} (a). $T_{2E}$ determined from the offset is approximately 18~$\mu$s. Measurements on LSCO-7$\%$ give roughly the same value. We thus conclude that the nontrivial behavior of the spin-spin relaxation rate originates in the Redfield term.

The same scaling can be attempted for measurements at different frequencies across the spectral line, as shown for LSCO-8$\%$ in Fig. \ref{t1t2frek} (b). Overall, the agreement is again fairly good. Interestingly, while the scaling near the center of the line is obeyed at all measured temperatures, the overlap at the line edges is not as convincing. This could imply that additional contributions to the relaxation mechanism appear in regions where strong spin fluctuations are present. 

\subsection{Temperature dependence of relaxation rates}

We find that the temperature dependence of the spin-spin relaxation rates is similar for all measured samples, up to a doping-dependent offset (Fig. \ref{T2_temp_svi} (b)). This has already been observed for spin-lattice relaxation in previous work on samples with higher doping levels (ref.  \citenum{Gorkov2004}), and we show here that the trend continues at least down to $x$ = 5$\%$. From the scaling described in the previous subsection, we have demonstrated that the temperature dependence of $T_2$ at the line center originates from changes in $T_1$ through the Redfield term. We can therefore use $T_2$ as a proxy for $T_1$, and compare the observed trends in $T_2$ with known $T_1$ results.  

The spin-lattice relaxation rates in LSCO are known to be significantly faster than in other cuprates, which show similar behavior both above and below $T_c$  \cite{Jurkutat2019, Avramovska2020}. The anomalous behavior of LSCO was attributed to an independent, parallel relaxation mechanism, possibly due to differences in the orbital makeup of the planar Cu and O atoms as compared to other cuprates. Moreover, an empirical model that has been proposed by Gor'kov and Teitel'baum (refs.  \citenum{Gorkov2004},  \citenum{Gorkov2005}) decomposes the Cu spin-lattice relaxation rates into two components: a universal temperature-dependent term, which is the same for all cuprates, and a term caused by spin stripes and disorder  \cite{Gorkov2004, Gorkov2005, Gorkov2006, Gorkov2015}.  The total relaxation rate is then
\begin{equation}
1/^{63}T_1=1/^{63}T_1(T)+1/^{63}T_1(x),
\end{equation}
where $T_1(T)$  only depends on temperature, while $T_1(x)$ is family- and doping-dependent. Vertical scaling of $1/T_1$ measurements for different samples has shown that results for all available doping levels collapse on a single curve for $x>$10$\%$ \cite{Gorkov2004}. Non-zero offsets appear in materials for which incommensurate peaks have been observed in neutron scattering, which have been interpreted as a signature of dynamic stripes or some other form of correlated electronic inhomogeneity. The remaining temperature dependence is nearly doping- and material-independent, and even appears in an electron-doped cuprate \cite{Gorkov2005elektroni}. This component was extensively investigated theoretically in several recent papers  \cite{Jurkutat2019, Avramovska2020}.

\begin{figure*}[t!]
\hspace*{0cm}
\includegraphics[trim={0cm 2cm 0cm 1cm},clip,  width=0.9\linewidth]{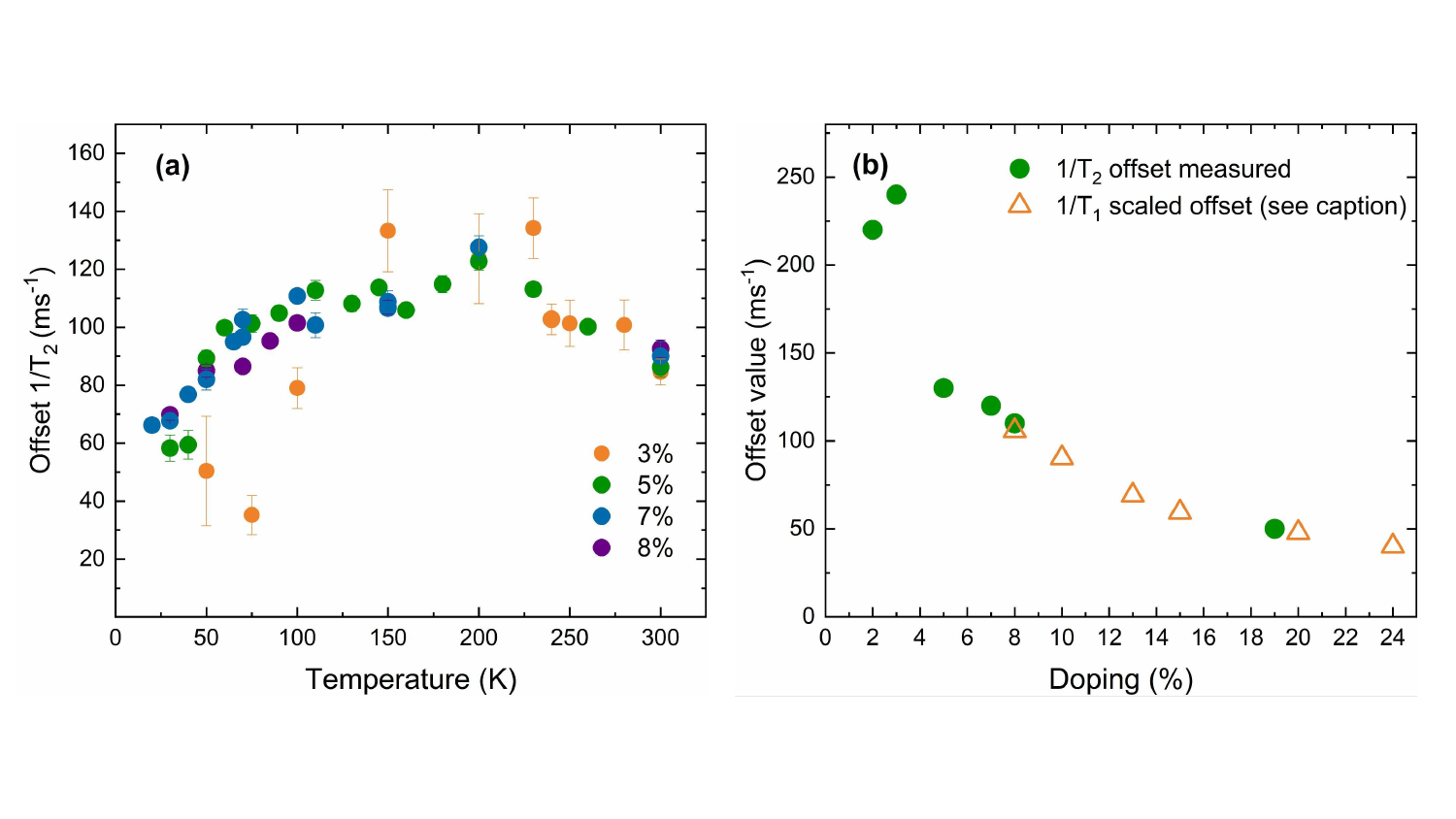}
\caption[Offsets for all measured samples] {\label{T2offset}\label{T2collapsed}(a) Temperature dependence of spin-spin relaxation rates offset by doping-dependent constants, chosen such that all data collapse on a single curve. (b) The offsets of measured 1/$T_2$ values, plotted together with scaled $T_1$ offsets determined by Gor'kov and Teitel'baum  \cite{Gorkov2004, Gorkov2006}. It can be seen that the  two datasets connect smoothly at lower doping, which suggests that the same considerations apply for $T_2$ as for $T_1$ .}
\end{figure*}

We expand the scaling analysis to the doping range studied here. The same basic relation seems to hold for our data:  $1/T_2$ values collapse to a single curve for all doping levels down to $x$ = 5$\%$ when corrected for temperature-independent offsets, as shown in Fig. \ref{T2collapsed} (a). The offsets increase monotonically with doping, similar to the $T_1$ offsets derived by Gorkov and Teitel'baum (ref.  \cite{Gorkov2004}). As a linear relation is expected between 1/$T_1$ and 1/$T_2$ offsets, and anisotropies are temperature-independent, the trends can be directly compared. Figure \ref{T2offset} (b) shows the values we obtain from $1/T_2$, together with previous $1/T_1$ results which we have linearly scaled to overlap with our data at two doping levels, $x$ = 19$\%$ and $x$ = 8$\%$. It can be seen that the $T_2$ offsets connect smoothly to the $T_1$ offsets, which suggests that the physical model of dynamical phase separation proposed by Gor'kov and Teitel'baum can be expanded down to $x$ = 5$\%$, but not below, consistent with a qualitative change occurring at this doping (Fig. \ref{cluster}). We emphasize that these conclusions pertain to the narrower NMR line component.

\subsection{Electronic inhomogeneity}

The non-monotonic relaxation rates across the spectral line that appear in our measurements at low temperatures are unusual, since they do not overlap at the same frequencies, indicating that they cannot be explained by simple doping heterogeneity. Qualitatively, our low-temperature results for $T_1$ and $T_2$ (Fig. \ref{T1frek}) are similar to the behavior observed, \textit{e.g.}, by Mounce \textit{et al.} (ref.  \citenum{Mounce2011}) for Bi-based cuprates in the mixed superconducting state. Their NMR study of superconducting vortex dynamics investigates the spatial distribution of local fields through the analysis of broad spectral lines. Because of the vortex structure, it is possible to clearly resolve regions in the vortex core from those that are far from it, as the varying magnetic field causes a distribution of resonance frequencies \cite{Sigmund2003, Mitrovic2001, Morr2000}. Moreover, similar NMR features have been related to the presence of charge order in LSCO-11.5$\%$. We therefore suggest that the broad NMR line components in samples with $x$ = 7$\%$ and 8$\%$ originate in local environments with enhanced antiferromagnetic fluctuations, in the vicinity of AFM ordered regions (see right panel in Fig. \ref{cluster}). These localized structures are characterized by different relaxation rates than those of the surrounding material. 
The geometry of these structures could be "droplets", or "patches" within the Cu-O planes, which evoke magnetic vortices, or disordered stripes, both of which have been suggested for cuprates such as LSCO  \cite{Andersen2010, Andersen2011,Christensen2011, Matsuda2000}. We emphasize that the broad NMR lines and stretched exponential relaxations are consistent with a wide distribution of the characteristic spin fluctuation scales, and thus do not directly support the presence of sharp phase boundaries between distinct electronic environments.

More broadly, our results provide microscopic evidence that electronic heterogeneity is ubiquitous in lightly doped LSCO, with the observation of a qualitative change around $x = 5$\%, the doping level where superconductivity appears in the phase diagram. We note that qualitatively similar conclusions have been obtained for Mott transitions in other materials such as nickelates, V$_2$O$_3$ and rare-earth titanates  \cite{Frandsen2016,Hameed2021}. We now discuss a simple real-space picture based on refs.  \citenum{Frandsen2016} and  \citenum{Hameed2021} (Fig. \ref{cluster}) to rationalize our NMR results and connect them to other probes, most notably transport. At the lowest doping levels and low temperatures, the doped holes form metallic nanoregions within the AFM ordered material (with the order not necessarily long-range). Dynamic AFM fluctuations are still strong in these metallic regions, and we thus observe wide spectral lines with fast nuclear relaxation, such as those measured in LSCO-2$\%$ and 3$\%$. Importantly, the NMR signals of nuclei within the AFM phase without itinerant charge are essentially unobservable, due to large local fields that induce extreme line broadening. In this regime, there is no coherent transport between the metallic islands, charge-glass behavior sets in at the lowest temperatures, and the resistivity is consistent with variable-range hopping  \cite{Ando2004, SeikiKomiya_2009}. With increasing temperature, the AFM correlations weaken, the metallic regions effectively expand, as seen from the decrease of the NMR wipeout fraction. As the doping level increases, the metallic regions grow, and conduction across the insulating regions at low temperatures can be established \textit{via} tunneling. This leads to the appearance of a logarithmic temperature dependence of the low-temperature normal-state resistivity  \cite{Ando1995} and an abrupt change in the NMR properties around $p = 5$\%. A new normal state is established, that shows no qualitative change upon further doping, either in transport or in NMR  \cite{Ando1995, Frachet2020}. 

In this scenario, the transition around $x$ = 5$\%$ is based on connectivity between metallic regions, and thus related to percolation. This is qualitatively consistent with observations of intrinsic inhomogenity and percolation physics in superconducting  \cite{Pelc2018,Pelc2019} and structural  \cite{Pelc2021} fluctuations in LSCO and other cuprates. Studies on LSCO films have obtained evidence for phase competition around $x$ = 5$\%$ as well, and our results provide a microscopic underpinning for these observations \cite{Li2016}. The transition is likely not directly related to the formation of Cooper pairs, as it appears irrespective of applied magnetic field. It thus seems plausible that the basic underlying phenomenon is an insulator-metal transition, and that superconducting pairing develops once low-temperature coherent transport is possible. 

The geometry of electronic heterogeneity in the doping range above $x$=5$\%$ remains an important open question, since the mesoscale real-space structure cannot be determined from NMR. Moreover, NMR is a low-frequency probe, and does not provide comprehensive insight into the spin dynamics. However, some qualitative conclusions can still be made. We have shown that the relaxation rates of the line center -- that likely corresponds to metallic regions -- can be described as a superposition of two components, in accordance with a dynamic phase separation model. This implies that spin fluctuations are still strong in the metallic regions, but with characteristic fluctuation timescales significantly shorter than the NMR frequency. In contrast, the fluctuations become slow or quasi-static in the insulating regions, which leads to NMR line broadening, short spin-spin relaxation times, and signal wipeout. As noted, the boundaries between metallic and insulating regions are likely not sharp, but instead involve a gradual slowing down of the spin fluctuations, which leads to a wide distribution of relaxation rates and broad NMR lines. It is therefore plausible that the main difference between metallic and insulating regions is the timescale of the spin fluctuations, \textit{i.e.} that the phase separation is intimately related to dynamical heterogeneity.

While LSCO is one of the most heavily studied cuprates, the system shows idiosyncrasies that make it difficult to gauge the importance of the observed phenomena for cuprate physics more broadly. It is known, for example, that La-based cuprates exhibit strong charge disorder due to substitutional doping  \cite{Singer2002}, along with different forms of octahedral tilt instabilities  \cite{latticerev,LSCOtilt}. Spin fluctuations are also generally stronger in LSCO than in other cuprate families, as seen in previous NMR work  \cite{Ohsugi1994}. As noted, several prominent cuprate families exhibit spin inhomogeneity and glassy behavior at low doping, but such effects have not yet been detected in materials such as HgBa$_2$CuO$_{4+\delta}$ (Hg1201), and it would be interesting to extend the present investigation to these systems. In this context, we note that qualitative differences in the spin excitation spectra have been found between underdoped Hg1201 and materials such as LSCO and YBCO: the latter exhibit the well-known hourglass dispersion with incommensurate low-energy fluctuations  \cite{hourglass}, while in Hg1201 the excitations are gapped down to the lowest available doping levels  \cite{Chan2016,Anderson2024}. Interestingly, the observed decrease of the spin-lattice relaxation rate (Fig. \ref{T1_fitovi}) is consistent with the presence of a gap in the spin excitation spectrum within the metallic regions. Therefore, it is a distinct possibility that, in the case of LSCO, the low-energy fluctuations observed with neutron scattering originate mostly from the insulating regions, while the gapped fluctuations are a general feature of the cuprates' metallic state.

\section{Summary}
To summarize, we have performed systematic Cu NMR measurements in crystals of lightly doped LSCO, in a region of the temperature-doping phase diagram that has until now been mostly unavailable to standard NMR techniques. Both the spectral lineshapes and shifts have been studied, as well as temperature- and frequency-dependent relaxation times $T_1$ and  $T_2$.

Our results show that a qualitative change occurs between $x$=3$\%$ and 5$\%$, coincident with the doping range where superconductivity appears in LSCO. For samples with $x<=3$\%, the detectable NMR signal around the central transition frequency is broad and rapidly relaxing, with the relaxation time $T_2 \sim$5$\mu$s nearly independent on frequency. The peak frequency is significantly lower than that extrapolated from results at higher doping levels, where the variation stems mostly from quadrupolar effects. For $x$= 5$\%$, the relaxation times are longer, typically $\sim$ 12$\mu$s, but exhibit a strong frequency dependence across the spectral line. A further increase in doping does not cause new qualitative changes in spectral width, position, or dynamics at high temperatures. In this regime, doping heterogeneity and quadrupolar shift account for variations in spectral position and relaxation rates across the spectral line, while $T_2$ variations are mostly caused by $T_1$ changes, and can be smoothly extrapolated from higher doping levels. These results are consistent with a low-temperature insulator-to-metal transition around $x$=5$\%$, when tunneling is established between insular metallic regions, and a granular metal is formed. 

Beyond $x$ = 5$\%$, we find evidence of growing electronic heterogenity low temperatures from the lineshapes and relaxation times, where two distinct contributions can be discerned -- a relatively narrow central component that likely corresponds to metallic regions, and a broad contribution with fast relaxation rates at the line edges. These "wings" are present for doping levels up to $\sim$ 11.5$\%$, but have not been observed below $x$ = 5$\%$. The latter could be due to the fast relaxation rates throughout the line and strong signal wipeout in this region of the phase diagram. Our results for LSCO-5$\%$ show that relaxation rates at the edge of the central transition line are faster than at the center even tough no visible "wing" signal is present.

Signal wipeout, which is present in all samples, is consistent with the appearance of regions with strong and relatively slow spin fluctuations. Unlike in other compounds, it is not a dynamic effect caused by a global increase in relaxation rates, at least within the timescales accessible to our setup. The fact that wipeout is more prominent for samples with lower doping levels indicates that the missing signal could be a measure of the volume fraction of the strongly fluctuating regions.

\acknowledgements{
We thank M. S. Grbi\'{c}, I. Jakovac and D. K. Sunko for comments and discussions, and W. Tabis for help with Laue sample orientation.
The work at the University of Zagreb was funded by the Croatian Science Foundation, Grant Nos. IP-2018-01-2970, UIP-2020-02-9494 and IP-2022-10-3382, and using equipment funded in part through project CeNIKS co-financed by the Croatian Government and the European Union through the European Regional Development Fund - Competitiveness and Cohesion Operational Programme (Grant No. KK.01.1.1.02.0013). The work at the University of Minnesota was funded by the US Department of Energy through the University of Minnesota Center for Quantum Materials, under Grant No. DE-SC-0016371. The work at TU Wien was supported by FWF Project P 35945-N. 
}

\newpage
%\bibliography{eg_bibliography}

\end{document}